\documentstyle[12pt,aps]{revtex}
\renewcommand{\baselinestretch}{1.25}
\newcommand{\be}{\begin{equation}}
\newcommand{\ee}{\end{equation}}
\newcommand{\bee}{\begin{eqnarray}}
\newcommand{\eee}{\end{eqnarray}}

\newcommand{\dj}{d \raisebox{1.4mm}{\hspace*{-2.75mm}-}
                                \hspace*{-1.5mm}}

\newcommand{\pieta}{$\pi^- p \rightarrow \eta n$\ }
\begin{document}
\begin{titlepage}
\vspace*{3.cm}
\begin{center}
      {\Large  \bf $\eta N$ S-wave scattering length in a three coupled
      channel, multiresonance, unitary model }
\vspace*{10mm}\\
Mijo Batini\'{c}, Ivo \v{S}laus, Alfred \v{S}varc   \\
{\em Ru\dj er Bo\v{s}kovi\'{c} Institute,  Zagreb,
 \vspace*{1.cm} Croatia }
\end{center}
      The  S-wave  scattering  length for $\eta N$ elastic scattering is
      extracted from the S-wave T-matrix in  a  three  coupled  channel,
      multiresonance  unitary  model.   Results are compared with values
      already  reported  in  literature  which  are  obtained   applying
      multichannel,  but  single  resonance  -- no background models.  A
      dispersion among the previously published values of the real  part
      of  the S-wave scattering length is observed.  We demonstrate that
      the reported spread originates from the strong sensitivity of  the
      scattering  length  upon  the  small  variation  of the used input
      resonance  parameters.   In  addition,  we  show  that  $\eta   N$
      scattering  length  value  obtained  in  single  resonance  --  no
      background models significantly increases if  background  term  is
      added in a unitary way.  We question the reliability of previously
      reported  values  based  only  on  the  single  resonance  --   no
      background  models, and demonstrate that the value of the $\eta N$
      S-wave scattering length obtained in this publication is much more
      realistic because of the multiresonance and unitary approach.
      \vspace*{0.2cm} \\ PACS number(s):  13.60.Le, 14.40.Aq
\end{titlepage}
\renewcommand{\baselinestretch}{1.8}
\newpage
\setcounter{page}{1}
      In 1985 Bhalerao and Liu \cite{Bha85} have constructed  a  coupled
      channel  isobar  model  for  the $\pi N \rightarrow \pi N$, $\pi N
      \rightarrow \eta N$ and $\eta N \rightarrow \eta N$ T  -  matrices
      with  $\pi N$, $\eta N$ and $\pi \Delta$ ($\pi \pi N$) as isobars.
      A single  resonance  separable  interaction  model  for  S$_{11}$,
      P$_{11}$, P$_{33}$ and D$_{13}$ partial waves has been used.  They
      have used only $\pi N$ elastic scattering  data  as  a  constraint
      while their prediction for the $\eta$ production cross section has
      been  compared  with,  at  that  time   the   most   recent   data
      \cite{Bro79}.   Their conclusion has been that the S-wave $\eta N$
      interaction is attractive, and they have extracted for the  S-wave
      scattering length the value of $a_{\eta N}=(0.27 + i \, 0.22)$ fm.

      Arima  et  al  \cite{Ari92}  have studied the nature of the S-wave
      resonances  S$_{11}$(1535)  and  S$_{11}$(1650)  concerning  their
      couplings with the $\eta N$ channel using the two quark-model wave
      functions with pure intrinsic spin states for  the  isobars.   The
      dynamical coupling of the isobars to $\pi N$ and $\eta N$ channels
      are  described  by  the  meson-quark  coupling.   In  addition  to
      analyzing  the agreement of the model with the $\pi N$ elastic and
      $\eta$ production data they have obtained  the  S-wave  scattering
      length $a_{\eta N} = (0.98 + i\,0.37)$ fm.

      Wilkin \cite{Wil93} based his calculation on an  S-wave  threshold
      enhancement  calculation, used the $\eta$ total cross section near
      threshold to fix the imaginary part of the T-matrix  and  obtained
      the real part by fitting the \pieta production cross section up to
      the center of  mass  momentum  in  the  $\eta  n$  system  of  1.2
      (fm$^{-1}$).   He quotes the value of $a_{\eta N} = (0.55 \pm 0.20
      + i\,0.30)$ fm.

      Abaev and Nefkens \cite{Aba94} have also used a form of  a  S-wave
      single  resonance  model,  adjusted  the  resonance  parameters to
      reproduce the \pieta production  channel  to  the  best  of  their
      ability  and  extracted the S-wave scattering length as:  $a_{\eta
      N} = (0.62 + i \, 0.30)$ fm.

      The  large  spread in these values of the fundamental $a_{\eta N}$
      parameter requires the need for better understanding of the  $\eta
      N$  system  at  low energies.  The first step to achive this is to
      obtain a reliable set of $\pi N \rightarrow \eta N$ T-matrices.

      The indication of strong and attractive $\eta N$  interaction  has
      led  to a speculation about the existence of a new type of nuclear
      matter -- quasi-bound $\eta$-mesic nuclei \cite{Liu86}.  The
      properties  of  this  new  matter  are  determined by the $\eta N$
      interaction at low energies.

      Good  data  on  $\eta$  production  in  $\pi^- p$  interaction are
      missing.  The dominant contribution to the surprisingly big $\eta$
      production  channel  is  coming from the S$_{11}$(1535) resonance.
      The  contributions of  the   P$_{11}$(1440)   and   D$_{13}$(1520)
      resonances are  important, but not completely clarified.  The role
      of other resonances, even in these partial waves, is  not  at  all
      discussed  because of the single resonance character of the models
      \cite{Bha85,Ari92,Wil93}.      Recently,      accurate      $\eta$
      photoproduction   data   have   been  obtained  by  TAPS  at  MAMI
      \cite{Kru93} up to $E_{\gamma}$ = 790 MeV.   These  data  indicate
      that the D$_{13}$ resonance contribution is small.

      Several  attempts to extract the $\eta N$ S-wave scattering length
      $a_{\eta N}$ on the basis of complete knowledge of $\pi N$ elastic
      channel  and  a  part of the $\eta$ production data have been made
      \cite{Bha85,Ari92,Wil93,Aba94,Liu93}  using  models  of  different
      complexity.    However,   with   the  exception  of  Arima  et  al
      \cite{Ari92} only a single resonance per  partial  wave  has  been
      used, therefore, producing little information about the importance
      of other resonances and background terms.  The imaginary parts  in
      all  analyses  have  shown  a  fair  amount  of agreement, but the
      outcome for the real parts differs drastically from case to  case,
      showing   the   spread  from  0.27  fm  \cite{Bha85}  to  0.98  fm
      \cite{Ari92}.

      We shall show that the complete (and not only  partial)  knowledge
      of  the  $\eta N$ S-wave T-matrix is essential in order to extract
      the  real  part  of  the  S-wave  scattering  length.   Therefore,
      attempts  not  founded  on a multichannel, multiresonance, unitary
      representation of the S-wave T-matrix should be  considered  as  a
      rough  estimate only.  The S-wave scattering length, coming out of
      our three coupled channel,  multiresonance  analyses  \cite{Bat94}
      gives  higher  values  for the scattering length magnitude than it
      has  been  previously  reported  in  all  publications  with   the
      exception  of  Arima  et.   al.   \cite{Ari92}.  All values of the
      extracted $a_{\eta N}$ are given in \vspace*{1.0cm} Table 1.  \\
{\large \bf \vspace*{1.0cm} Formalism} \\
      \indent
      The  formalism  used  in  this  article  is  a  standard  textbook
      formalism  for  extracting  the  scattering  length from the known
      partial wave T-matrix.  We give here the  essential  formulae  for
      the  convenience  of  the  reader.   The  partial wave S matrix is
      written in the following way:
      \be
      S_{l}=\frac{1+i \ k_{l}(p)}{1-i \ k_{l}(p)} \ \ ; \ \
                    k_{l}(p)=\tan \delta_{l}(p),
      \ee
      $p$  is  center  of  mass  momentum,  $l$  is the angular momentum
      quantum number and $\delta$ is the partial wave phase  shift.   \\
      We  use the low energy expansion\footnote{Different textbooks take
      opposite sign for  the  scattering  length,  see  \cite{Gol64}  vs
      \cite{Tay72}.     \vspace*{-0.3cm}    Here    we    follow    Ref.
      \cite{Gol64}.}:  \\
      \be
      p^{2l+1} \cot \delta_{l} = \frac{1}{a_{l}} + \frac{r_{l}}{2} \
                 p^{2} + O(p^{4})
      \ee
      where $a_{l}$ and $r_{l}$ are scattering length and effective
      range, respectively. \\
      Using the definition of the partial wave S-matrix for the S-wave
      we obtain  the expression for the scattering amplitude $f_0(p)$:
      \be
      f_{0}(p)=\frac{T_{0}}{p}=\frac{1}{2ip}[S_{0}(p) - 1] \approx
      \frac{a_{0}}{1 - ia_{0}p + \frac{1}{2} a_{0} r_{0}p^{2} }
      \label{eq:sampl}
      \ee
      So  the  final  expression  for the S-wave scattering length $a_0$
      used through this article is given as:
      \be
          a_{0} = \lim_{p \rightarrow 0}
               \frac{f_{0}(p)}{1+ i \  p f_{0}(p) - \frac{1}{2} f_{0}(p)
              r_{0} p^{2} }
      \label{eq:a0}
      \ee
      and can be approximated by
      \be
          a_{0} \approx \frac{f_{0}(p)}{1+ i \  p f_{0}(p)}
      \ee
      for  the   small   $p$.    The   $\lim_{p   \rightarrow   0}$   in
      Eq.(\ref{eq:a0}) is done numerically.  \\
      When  only the elastic channel is opened, the scattering length is
      a real quantity.  Upon  opening  other,  inelastic  channels,  the
      scattering  length  becomes  complex  and  its  imaginary  part is
      related to the total cross sections via the optical theorem.   For
      the multichannel case \cite{Gol64,Tay72} we have
\be
      Im\ f_{\alpha \rightarrow \alpha}(p_\alpha,\vartheta=0)
      = \frac{p_\alpha}{4\pi}
      \sum_{x}\sigma^{tot}_{\alpha \rightarrow x}
\ee
      where  $x$  denotes  all  opened  channels.   In our model we have
      decided to take, in addition to two  physical  two  body  channels
      $\pi  N$  and  $\eta  N$,  an effective two body channel $\pi^2 N$
      which collects all remaining two and many body contributions.   In
      case  of  s  wave  scattering  at $\eta N$ threshold ($f_0 \approx
      a_0$), we get
\be
      Im\ a_0(\eta N) = \frac{1}{4\pi} \lim_{p_\eta \rightarrow 0}
      p_\eta \left(
                  \sigma^{tot}_{\eta N \rightarrow \pi N}
                + \sigma^{tot}_{\eta N \rightarrow \pi^2 N}
                \right)
\ee
      where $p_\eta$ is the c.m.  momentum of the particles in the $\eta
      N$  channel.   Note  that  the  contribution  of  $\eta N$ elastic
      scattering  is  zero  because  $\lim_{p_\eta  \rightarrow   0}   \
      p_\eta\sigma^{tot}_{\eta  N  \rightarrow  \eta  N}  = 0$.  For the
      other  two  channels  the  $p_{\eta}$  factor   appears   in   the
      denominator  of the expression for the total cross section because
      of  the  input  flux  factor,  and  that  leads  to   the   finite
      contribution to the imaginary part of the scattering length.

      Using isospin algebra and detailed balance we get  the lower bound
      for the imaginary part of the S-wave scattering length:
\be
      Im\ a_0(\eta N) \ge \frac{3p_\pi^2}{8\pi}
      \frac{\sigma^{tot}_{\pi^- p \rightarrow \eta n}}{p_\eta},
\ee
      where the $p_\pi$ is the c.m.  momentum of the  particles  in  the
      $\pi  N$  channel  at  $\eta N$ threshold.  Using the experimental
      value
\bee
      \frac{\sigma^{tot}_{\pi^- p \rightarrow \eta n}}{p_\eta}
      &=& (21.2 \pm 1.8)\ \mu{\rm b} / {\rm MeV}
\label{eq:XStot}
\eee
      from Ref. {\rm \cite{Bin73}}  we obtain the optical theorem
      constraint  based on the experimental $\sigma_{tot}(\pi N
      \rightarrow \eta N)$ value:
\be
      Im\ a_0(\eta N) \ge (0.24 \pm 0.02)\ {\rm fm}.
\label{eq:aEst}
\ee
      Keeping  in  mind  that  for  the  lowest  $S_{11}$  resonance the
      branching ratio to $\pi^2 N$ channel ($\pi\  N$,  $\pi\Delta$  and
      others)  is  5--20  \% \cite{Pdg92}, we make an estimate that $Im\
      a_0(\eta N) \approx 0.30$ fm.  However, let us  mention  that  our
      analysis \cite{Bat94} prefers somewhat smaller branching ratio, we
      get  3  $\pm$  3  \%  which  agrees with  the  value   quoted   in
      \cite{Pdg94}.   The  identical  method  of  using the unitarity to
      extract the lower bound for the imaginary part  of  the  $\eta  N$
      scattering   length   has   been   used   in  Ref.   \cite{Wil93}.
      \vspace*{1.0cm} \\
      {\large  \bf  Previously reported S-wave scattering length
      \vspace*{1.0cm} values} \\
      \indent
      The $\eta N$ S-wave scattering length has been extracted from $\pi
      N$  elastic  data  using  different  forms  of multichannel single
      resonance models with number of  channels  reduced  to  two.   The
      constraint  to  the $\pi N \rightarrow \eta N$ total cross section
      is imposed in \cite{Wil93}, while other authors have only compared
      the  outcome  of their analysis with $\eta$ production total cross
      section   \cite{Ari92,Wil93,Aba94}.    Only   authors   in    Ref.
      \cite{Bha85}  compare  their  result  with  the  $\eta$ production
      differential cross section measurements.  We collect all  reported
      values \cite{Bha85,Ari92,Wil93,Aba94,Liu93} in Table 1.

      All  extracted values for the imaginary part agree reasonably well
      in spite  of  the  differences  of  the  used  models  (previously
      described).   This  is  a  direct consequence of the fact that the
      optical theorem via unitarity is build into each of them.  Let  us
      point  out  that  only  one  analysis  \cite{Bha85}  violates  the
      experimental optical theorem value given in Eq.   (\ref{eq:aEst}),
      and  we will now explain why,  and how it can be modified.  In the
      model of Ref.  \cite{Bha85} the $\pi  N$  elastic  scattering  has
      been  used  to  constrain  all  free  channel  parameters  of  the
      analysis, including $\eta  N$  channel.   Therefore,  the  $\pi  N
      \rightarrow  \eta  N$  differential cross section is a prediction.
      The obtained predicted values  have  been  compared  only  to  the
      experimental  data  of  Ref.   \cite{Bro79} at energies up to 1572
      MeV.  These data tend to be much lower than the results  of  other
      measurements.  Since data of Ref.  \cite{Bro79} are now considered
      to be questionable, there  is  a  need  to  correct  the  obtained
      scattering length value in order to improve the agreement with the
      recently recommended \cite{Cla92} \pieta data.

      A  simple  estimate  of  how  much  a scattering length, quoted in
      \cite{Bha85}, will  change  if  the  condition  to  reproduce  the
      recommended  \cite{Cla92} \pieta cross sections is imposed, can be
      done in the following way:

      If  a  restriction of using only one resonance per partial wave is
      imposed (and \cite{Bha85} is a single resonance model), the  $\eta
      N$  total  production cross section very close to the threshold is
      given by:
      \be
      \frac{\sigma^{tot}_{\pi^{-} p \rightarrow \eta n}}{p_{\eta}}
      \approx \frac{2}{3} \ \frac{4 \pi}{p_{\pi}^2} \ |T_{\pi \pi}| \
      |a_{\eta N}|.
      \ee
      This gives the values of 15.0 $\mu$b/MeV and 13.4  $\mu$b/MeV  for
      each  of the two solutions of Ref.  \cite{Bha85} respectively.  If
      a value for the $\pi N$ elastic S-wave T-matrix at  threshold  are
      read  of  the graph from Ref.  \cite{Bha85}, and are taken to be :
      $T_{\pi \pi}^1=(0.38 + i \, 0.31)$ and $T_{\pi \pi}^2=(0.37 + i \,
      0.25)$  the  correction factors $M_{j}$;  j=1,2 needed to obtain
      the   recommended   cross   section   values    given    in    Eq.
      (\ref{eq:XStot}) turn out to be:
      \be
           M_{j} =
      \left\{  \begin{array}{ll}  1.41  &  \mbox{{\rm  \  for  \  the  \
                          solution} \ (0.27+ i \, 0.22); \ \ j=1} \\
        1.58 & \mbox{{\rm \ for \ the \ solution} \ (0.28+ i \, 0.19); \ \ j=2
}
                    \end{array}
                  \right.
      \ee
      The resulting "modified" $\eta N$ scattering length solutions are:
      \bee
        (0.27+ i \,0.22) \ {\rm fm} & \longrightarrow & (0.38 + i \, 0.31) \
{\rm fm} \nonumber \\
        (0.28+ i \,0.19) \ {\rm fm} & \longrightarrow & (0.44 + i \, 0.30) \
{\rm fm}
      \eee
       Therefore, if a consistent fit of \pieta data is used, the outcome
       for  imaginary  part  of  the  $\eta  N$  scattering length in Ref
       \cite{Bha85} becomes  consistent  with  the  value  given  in  Eq.
       (\ref{eq:aEst}).

      Real  part  of  the  $a_{\eta N}$, however, shows a notable spread
      among the models ( 0.27 $ \leq {\rm Real}(a_{\eta N}) \leq $  0.98
      )  in  spite of the fact that almost identical data have been used
      as input.  The origin of this disagreement has not been identified
      up  to  now.   The  aim of this paper is to address and to explain
      this disagreement within the framework of single resonance models,
      and  to  give the value for the $a_{\eta N}$ when the extension to
      multiresonance unitary models with background explicitly  included
      is \vspace*{1.0cm} done.  \\
{\large \bf Single S-wave resonance model \vspace*{1.0cm} (SR)} \\
      \indent
      The single resonance model (SR) can be introduced in at least two
      ways:
      \begin{description}
      \item[a.]  to  use  the  resonance  parameters  \cite{Pdg92,Pdg94}
      directly
      \item[b.] to simulate a single resonance model within the scope of
      three coupled channel, multiresonance, unitary model \cite{Bat94}
      \end{description}

      We have tested both approaches.  As it  is  to  be  expected,  the
      outcome is very similar.

      We have tested the behavior of the $\eta N$ scattering length with
      respect to the probable uncertainty of resonance parameters  in  a
      single  resonance  model.   We  have used the resonance parameters
      \cite{Pdg92,Pdg94}, and we have allowed for their variation in the
      following way:
\bee
      M_{R}  & = & ( 1535 \pm 10) \ {\rm MeV} \nonumber \\
      \Gamma & = & (\ 150 \pm 20) \ {\rm MeV} \nonumber \\
\label{eq:ResPar}
      x_{\pi}& = & (\ 0.4 \pm 0.05)
\eee
      Instead of using the $\eta N$ branching ratio $x_\eta$ explicitly,
      we have used the fact that it is proportional to $\sigma_{tot}(\pi
      N \rightarrow \eta N)$ in all single resonance models, and we have
      taken a directly measured value of  total  cross  section  as  the
      input  parameter.   As  an  illustration,  the  value given in Eq.
      (\ref{eq:XStot}) for the $\pi^- p \rightarrow \eta n$ total  cross
      section together with values from Eq.  (\ref{eq:ResPar}) gives
\be
      x_\eta = 0.405 \pm 0.023.
\ee
      Note that in \cite{Pdg92} only a band of values\footnote{ Citation
      from  \cite{Pdg92}  follows:   Resonance  mass  M  =  1520 to 1555
      \vspace*{-0.3cm} ($\approx 1535$) MeV, full width $\Gamma$  =  100
      to  250  ($\approx  150$)  MeV,  fraction  $\Gamma_i$/ $\Gamma$ is
      35--50 \% for $\pi N$, 30--50 \% for $\eta N$ \vspace*{-0.3cm} and
      5--20  \% for $\pi \pi N$ channel.  Let us still point out that in
      Ref \cite{Pdg94} the  suggested  value  \vspace*{-0.3cm}  for  the
      $\pi^2$  branching  ratio  is  only  1  -- 10 \%.} instead of real
      statistical error is given  for  all  resonance  parameters.   The
      chosen  variations  in  Eqn.  (\ref{eq:ResPar}) are, therefore, an
      expression  of  our  decision,  and  are  intended  to  show   the
      sensitivity  of scattering length to a relatively small changes of
      input values.  Of course,  our  choice  is  within  the  suggested
      bands.

      Using  the  Cutkosky's  unitary  formalism  \cite{Cut79}  we  have
      obtained  the  S-wave T-matrix in the single resonance model using
      the parameter values from  (\ref{eq:ResPar}).   The  corresponding
      S-wave  scattering  length  is  obtained as the $p=0$ value of the
      scattering amplitude $f_0(p)$ defined  in  Eq.   (\ref{eq:sampl}).
      The   numerical   value   of   the   scattering  length  with  the
      uncertainties coming from the allowed variations of the  resonance
      parameters is given as follows:
\begin{center}
\begin{tabular}{clllll}
 $a_{\eta N}$&$ = $& $(0.404$ & $\ +\ i\ 0.343)$\ fm \\
   \hline
      &$\pm$&(0.085&$\  +\ i\ 0.046)$\ fm&\ \ using \ \ $\pm 10 $\ MeV &
      \ \ for $M$ \\
      &$\pm$&(0.053&$\ +\ i\ 0.020)$\ fm&\ \ using \ \ $\pm 20 $\ MeV  &
      \ \ for $\Gamma$ \\
      &$\pm$&(0.050&$\  +\ i\ 0.023)$\ fm&\ \ using \ \ $\pm 0.05$ & \ \
      for $x_\pi$ \\
      &$\pm$&(0.034&$\ +\ i\ 0.018)$\ fm&\ \  using  \  \  $\pm  1.8  $\
      $\mu$b/MeV  &  \  \  for $\frac{\sigma_{tot}(\pi^{-} p \rightarrow
      \eta n)}{p_{\eta}}$\\
   \hline
  &$\pm$&(0.117&$\ +\ i\ 0.058)$\ fm&\ \ \ Total
\end{tabular}
\end{center}
      (see Table 1.).

      The  dependence  of  $\eta  N$ scattering length on $M$, $\Gamma$,
      $x_\pi$  and  $\frac{\sigma_{tot}(\pi^{-}   p   \rightarrow   \eta
      n)}{p_{\eta}}$  is  shown  on  Figures~1a--1d.   One  parameter is
      variable, and other three are fixed to the values  given  in  Eqs.
      (9)  and  (\ref{eq:ResPar}).   Error  bands on Figures~1a--1d were
      obtained using the standard statistical definition  of  the  total
      error   of   the  function  which  depends  on  several  uncertain
      parameters.  Upon closer inspection of Figures~1a--1d we  conclude
      that  the  small variation of input parameters causes a big change
      in the resulting scattering length.  We conclude that the  present
      level   of   confidence   of  S$_{11}$  resonance  parameters  and
      constraint on $\pi N \rightarrow \eta N$ cross  section,  see  Eq.
      \ref{eq:XStot},   is  not  sufficient  to  predict  the  $\eta  N$
      scattering length to a level better than 50 \%.

      Fig~2a shows the comparison of the obtained single resonance  $\pi
      N$  elastic  T-matrix,  corresponding to the parameter values from
      Eqs.  (9) and (\ref{eq:ResPar}), with the partial wave analyses of
      H\"{o}hler  et.   al.  \cite{Hoe83}.  As the K-H PWA does not give
      the  error  analysis  for   the   partial   wave   T-matrices   in
      \cite{Hoe83},   and   the  errors  are  essential  to  define  the
      statistical weight of the analyses, we have identified the  errors
      of the used data in the standard $\chi^2$ analysis as:
      \begin{eqnarray*}
      \Delta_i &=& 0.005 + \left( 0.01 + 0.0015 \  \frac{W_i-W^{\pi
      \ {\rm thresh}} }
        { \Delta} \right) |T_{\rm max}| \\
      \Delta  &=&  1  \ {\rm GeV} \\ W_i & {\rm is} & {\rm the \ total \
      c.m.  \ energy } \\ W^{\pi \ {\rm thresh}}& {\rm is} & {\rm the  \
      total \ energy \ at \ \pi \ nucleon \ threshold} \\
             |T_{\rm max}| & {\rm is} &
              {\rm the \ maximal \ value \ of \ the \ S_{11}\ T-matrix \
              in \ the \ chosen \ energy \ range.}
      \end{eqnarray*}
      The energy range extents up to 2.5 GeV.

      The statistical weight in the $\chi^2$ function is defined in a
      standard way:
      \begin{eqnarray*}
          w_i=\frac{1}{(\Delta_i)^2}.
      \end{eqnarray*}

      The introduced energy dependence  of  the  statistical  weight  is
      inspired  by  the  energy dependence of the error analysis of Ref.
      \cite{Cut79}.  It steadily  raises  with  energy,  but  does  not
      exceed the value of 0.02 in the units of Ref. \cite{Hoe83}.

      The scattering amplitude in the SR model is  given  in  Fig.~3 and
      denoted  with the dotted line, the extrapolation to the $p_\eta=0$
      value (the S-wave scattering length)  is  marked  with  the  empty
      triangle  on  the  y-axes.

      The   agreement  between  predictions  for  the  $\pi  N$  elastic
      T-matrices, within the scope of {\em any} single resonance  model,
      and  the  standard  multiresonance  $\pi  N$ elastic input of Ref.
      \cite{Hoe83} is acceptable only in the vicinity of  S$_{11}$(1535)
      dominance.  That indicates problems occurring in any S-wave single
      resonance model in reproducing the "experimental" $\pi N$  elastic
      scattering   length\footnote{   As  it  is  pointed  out  in  Ref.
      \cite{Hoe83}  \vspace*{-0.3cm}  at  least  some  form   of   model
      dependence  have  to be introduced in extrapolating the scattering
      amplitude to the $\pi N$ elastic threshold.  \vspace*{-0.3cm}  So,
      there  can  exist  no  such  quantity as "experimental" scattering
      length.} because of the disagreement of  the  obtained  T-matrices
      with  well  known  input near $\pi$ threshold.  To demonstrate the
      problem we compare the "experimental" \cite{Hoe83}  and  SR  model
      values for the $\pi N$ elastic S-wave scattering lengths:
      \begin{center}
              $a_{\pi N}^{exp} = (0.249 \pm 0.004)$ fm \ \ {\rm and}
         \ \ $a_{\pi N}^{SR \ mod} =  0.066  $ \ fm.
      \end{center}
      The discrepancy is obvious. \\
      {\bf Conclusion:} {\em The result of our SR model is, as it is  to
      be  expected,  consistent  with  the  results  of all other single
      resonance models  \cite{Bha85,Wil93,Aba94}.   A  single  resonance
      model  which  is  based on Particle data group data for the lowest
      S-wave  resonance  \cite{Pdg92},  and  constrained   with   $\eta$
      production  total  cross  section \cite{Bin73} can not be reliably
      used for extracting the $\eta N$ S-wave scattering length.  It  is
      extremely   sensitive   to   the   precision  of  resonance  input
      parameters, and has notorious problems in failing to reproduce the
      $\pi  N$  elastic  scattering  length.   The  simplest improvement
      consists in including the background term in addition to a single
      \vspace*{1.0cm} resonance.} \\
      {\large \bf Single S-wave resonance + one  background  term  model
      \vspace*{1.0cm} (SRBG)} \\
      \indent
      To  improve the model and to manifestly demonstrate the importance
      of additional ingredients we have introduced one  background  term
      in  addition  to  the  SR  model.   The  $\pi  N$  elastic  S-wave
      scattering length, and high energy behavior  of  $\pi  N$  elastic
      S-wave  T-matrix  can  not  be  even  remotely  described  without
      introducing the background into the $\pi N$ elastic  channel.   Of
      course,  in  a  unitary model any modification in one channel will
      influence other channels as well.  So, consequently, the  addition
      of  the background term to the $\eta N$ channel is needed when the
      background  term  is  added  to  the  $\pi  N$  elastic   channel.
      Therefore,  changes  in  the  S-wave  scattering  lengths  in both
      channels are to be expected.

      We have used the formalism described in \cite{Bat94}.  However, as
      we are interested in the S-wave only, we perform here a fit of the
      background  term  parameters  to  the  $S_{11}$  $\pi  N$  elastic
      $T$-matrix \cite{Hoe83} up to the total c.m.  energy of  1560  MeV
      where  the  importance  of $S_{11}$(1535) resonance diminishes.  A
      constraint to the $\eta N$ channel is not yet imposed.  Note  that
      because of the formalism used, releasing the background parameters
      in  the  fitting  procedure  also   slightly   changes   resonance
      parameters.  The obtained parameters are:
      \begin{eqnarray}
      M^{res}&=&1538  \ {\rm MeV} \nonumber  \\
      \Gamma^{res}&=&127 \ {\rm MeV} \nonumber          \\
      \label{eq:ResSRBG}
      x^{res}_{\pi}&=&0.33           \\
      x^{res}_{\eta}&=&0.49  \nonumber         \\
      \frac{\sigma_{tot}(\pi  N  \rightarrow  \eta N)}{p_{\eta}}&=&9.2 \
      \mu {\rm b/MeV} \nonumber
      \end{eqnarray}
      Resulting T-matrix (full curves) is  compared  to  H\"{o}hler  PWA
      (full  dots)  \cite{Hoe83}  in  Fig.~2b.  Dashed and dotted curves
      represent   the   resonance    and    background    contributions,
      respectively.

      The  corresponding  S-wave  scattering  length  is obtained as the
      $p=0$ value of the scattering amplitude $f_0(p)$  defined  in  Eq.
      (\ref{eq:sampl}).   The  scattering amplitude in the SRBG model is
      given  in  Fig.~3  and  denoted  with   the   dashed   line,   the
      extrapolation  to  the  $p_\eta=0$  value  (the  S-wave scattering
      length) is marked with the empty inverse triangle on  the  y-axes.
      The  numerical  value  of the scattering length resulting from the
      resonance parameters of Eq.  (\ref{eq:ResSRBG}) is given by:
      \begin{eqnarray*}
      a_{\eta N}=(0.691 + i \ 0.174) \  {\rm fm}
      \end{eqnarray*}
      (see Table 1.).  As the SRBG model  is  a  simplification  of  our
      full,  multiresonance  model  \cite{Bat94}, and is given only as a
      demonstration of importance of different parts of the model.   For
      simplicity,  the error analysis is not given for SRBG.  However it
      is identical to the error analysis of our full model presented  in
      Ref.   \cite{Bat94},  and will be given in extracting the $\eta N$
      S-wave scattering length, in the final step, within the  scope  of
      our final CCMRU model.

      Real  part  of  the scattering length shows the strong tendency of
      rising ( 0.398 in SR vs 0.685 in  SRBG).   It  is  interesting  to
      point out that fitting only the elastic $\pi N$ part {\em strongly
      reduces the total cross section of $\eta$ production :  21.2 in SR
      vs  9.2  in  SRBG},  producing  the  analogous  reduction  of the
      scattering length imaginary part {\em $Im\ a_{\eta N}$ :  0.343 fm
      in SR vs 0.174 fm in SRBG.}

      The  agreement  of the SRBG model with the standard multiresonance
      $\pi N$ elastic T-matrix of Ref.  \cite{Hoe83} is now much  better
      near  the $\pi N$ threshold.  Therefore, the agreement of the $\pi
      N$ elastic scattering  length  within  the  SRBG  model  with  the
      "experimental"  value is much better.  To illustrate the effect we
      compare  the  "experimental" and SRBG model values for the $\pi N$
      elastic S-wave scattering lengths:
      \begin{center}
              $a_{\pi N}^{exp} = (0.249 \pm 0.004)$\ fm \ \  {\rm and}
       \ \      $a_{\pi N}^{SRBG \ mod} = 0.259 $\ fm.
      \end{center}
      However, the result for  $\sigma_{tot}(\pi^-  p  \rightarrow  \eta
      n)$,  obtained in the SRBG model, is not acceptable.  It completly
      disagrees  with  the   experimental   results.    Therefore,   the
      unavoidable  next  step  is releasing the resonance and background
      parameters in the fitting procedure imposing a constraint  on  the
      $\eta$  production  cross  section  at  the  same  time. \\
      {\bf  Conclusion:}  {\em  Introducing background term modifies the
      $\eta N$ scattering length significantly and the $\pi  N$  elastic
      scattering  length  is improved.  However, the free fit to $\pi N$
      data without any constraint to  the  $\eta  N$  channel  does  not
      reproduce  the  $\eta$  production  total  cross  section  at all.
      Therefore, a constraint of the fit with the $\eta N$ channel  data
      is \vspace*{1.0cm} needed.} \\
      {\large  \bf  Constrained  single resonance + one background
      model \vspace*{1.0cm} (CSRBG)} \\
      \indent
      Again,  we  have  used  the  coupled  channel,  multiresonance and
      unitary formalism presented in Ref.  \cite{Bat94}.  The  resonance
      and the background parameters have been simultaneously released in
      the fitting procedure, and we have fitted  the  S$_{11}$  $\pi  N$
      elastic S-wave T-matrix of Ref.  \cite{Hoe83} up to the total c.m.
      energy of 1560 MeV.   The  $\eta  N$  channel  is  constrained  by
      forcing  the  total $\pi N \rightarrow \eta N$ total cross section
      at threshold to agree with the experimental value.  That has  been
      done  in  a  standard  way  introducing  a  penalty  function into
      $\chi^2$.

      Resulting $\pi N$ elastic S-wave T-matrix (full curve) is compared
      to  H\"{o}hler's  PWA  (full  dots)  in  Fig.~2c  and the $\eta N$
      scattering length is given in Table 1.  Dashed and  dotted  curves
      represent    the    resonance    and   background   contributions,
      respectively.  The obtained resonance parameters have not  changed
      drastically except, of course, $x{_\eta}$:
      \begin{eqnarray}
      M^{res}&=&1537  \ {\rm MeV}   \nonumber        \\
      \Gamma^{res}&=&145 \ {\rm MeV}  \nonumber         \\
      \label{eq:ResCSRBG}
      x^{res}_{\pi}&=&0.29           \\
      x^{res}_{\eta}&=&0.70      \nonumber     \\
      \frac{\sigma_{tot}(\pi  N  \rightarrow \eta N)}{p_{\eta}}&=&21.2 \
      \mu{\rm b/MeV} \nonumber
      \end{eqnarray}
      The corresponding S-wave scattering length is obtained identically
      as before, as the $p=0$ value of the scattering amplitude $f_0(p)$
      defined in Eq.  (\ref{eq:sampl}).  The scattering amplitude in the
      CSRBG  model  is  given in Fig.~3 and denoted with the dash-dotted
      line, the  extrapolation  to  the  $p_\eta=0$  value  (the  S-wave
      scattering  length) is marked with the empty square on the y-axes.
      The numerical value of the scattering length  resulting  from  the
      resonance parameters of Eq.  (\ref{eq:ResCSRBG}) is given by:
      \begin{eqnarray*}
      a_{\eta N}=(0.968 + i \ 0.281) \  {\rm fm}
      \end{eqnarray*}
      (see Table 1.).
      The CSRBG model is a simplification of our full, multiresonance
      model \cite{Bat94}, and is given only as a demonstration of
      importance of different parts of the model.
      \footnote{The error analysis is not given for the same reasons as
      for SRBG model.}

      Because  of better agreement between the CSRBG model and the input
      $\pi N$ elastic S-wave T-matrices,  the  value  for  the  $\pi  N$
      elastic S-wave scattering length is improved.
      \begin{center}
              $a_{\pi N}^{exp} = (0.249 \pm 0.004)$\ fm\ \ and
         \ \ $a_{\pi N}^{CSRBG \ mod} = 0.251 $\ fm.
      \end{center}
      {\bf  Conclusion:}  {\em  Introducing  the  background  term  in a
      unitary way, and reproducing all experimental inputs,  shifts  the
      real  part of the scattering length to higher values.  The $\pi N$
      elastic scattering length is closer to the  "experimental"  value.
      The  agreement  of  the elastic $\pi N$ T-matrix with data of Ref.
      \cite{Hoe83} is not yet perfect, so the inclusion of other  S-wave
      resonances and background terms is \vspace*{1.0cm} needed.  } \\
 {\large \bf Three coupled channel, multiresonance and unitary model
     \vspace*{1.0cm} (CCMRU) } \\
     \indent
      We  have  constructed  a three coupled channel, multiresonance and
      unitary model (CCMRU) \cite{Bat94}, and fitted it to the  $\pi  N$
      elastic  partial wave T-matrices of Ref.  \cite{Hoe83} in 8 lowest
      $I=1/2$ partial waves:  S$_{11}$,  P$_{11}$,  P$_{13}$,  D$_{13}$,
      D$_{15}$,  F$_{15}$,  F$_{17}$  and  G$_{17}$,  up  to  total c.m.
      energy of 2500 MeV, and to the available $\pi^- p \rightarrow \eta
      n$  total  and  differential  cross  sections.  The results of the
      analysis are presented in Ref.  \cite{Bat94}, with  all  resonance
      parameters  explicitly  given therein.  For the convenience of the
      reader, the comparision of the S$_{11}$ $\pi N$  elastic  T-matrix
      with the input data of \cite{Hoe83} is given in Fig.~2d.

      The corresponding S-wave scattering length is obtained identically
      as before, as the $p=0$ value of the scattering amplitude $f_0(p)$
      defined in Eq.  (\ref{eq:sampl}).  The scattering amplitude in the
      CCMRU model is given in Fig.~3 and denoted with  the  full  curve,
      the  extrapolation  to the $p_\eta=0$ value (the S-wave scattering
      length)  is  marked  with  the  empty  circle  on the y-axes.  The
      numerical value of the scattering length, with the error  analysis
      given in Ref.  \cite{Bat94}, is given by:
\begin{center}
\begin{tabular}{ccc}
 3 resonances in $P_{11}$ partial wave: \
&\ [\ 0.886 $\pm$ 0.047\ &\ + i\ (0.274 $\pm$ 0.039)\ ] fm\\
 4 resonances in $P_{11}$ partial wave: \
&\ [\ 0.876 $\pm$ 0.047\ &\ + i\ (0.274 $\pm$ 0.039)\ ] fm
\end{tabular}
\end{center}
      and  we  recommend  it  as  being  more  reliable  then the values
      obtained by previous analyses.  The obtained  $\pi  N$  scattering
      length is in complete agreement with the "experimental" value:
\begin{center}
\begin{tabular}{cc}
$a_{\pi N}^{exp} = (0.249 \pm 0.004)$  fm   & \\
       and &   \\
 3 resonances in $P_{11}$ partial wave:
& \ \ $a_{\pi N}^{CCMRU \ mod}$= (0.247 $\pm$ 0.006)  fm \\
 4 resonances in $P_{11}$ partial wave:
& \ \ $a_{\pi N}^{CCMRU \ mod}$ = (0.248 $\pm$ 0.006)    fm
\end{tabular}
\end{center}
 {\large \bf Conclusion:} \\
      \indent
      We  do  not consider our result to be the final one because of its
      model dependence and because of the insufficient $\eta  N$  input.
      However,  it  is  certainly more reliable then the values reported
      within the scope of previous models.  The  new  experiments   near
      threshold  $\eta  N$ data are badly needed.  Proposed \cite{AGS94}
      measurements of the total an the differential  cross  section  for
      the  reaction  \pieta  from  threshold ($p_\pi = 685$ MeV/c) up to
      $p_\pi = 760$ MeV/c will result in a better experimental basis for
      the  reliable  extraction  of the $\eta N$ scattering length.  The
      accurate determination of the S-wave $\eta  N$  scattering  length
      will  shed   light  on  the  possible  existance  of a new kind of
      hadronic bound  state.   Namely,  the  indication  of  strong  and
      attractive $\eta N$ interaction has led to a speculation about the
      existence  of  a  new  type   of   nuclear   matter,   quasi-bound
      $\eta$-mesic  nuclei  \cite{Liu86}.   The  properties  of this new
      matter are determined by the $\eta N$ interaction at low energies.

      \newpage
      {\large \bf\vspace*{1.0cm}  Acknowledgment:} \\
      \indent
      We  are grateful to Prof.  B.M.K.  Nefkens and V.V.  Abaev for the
      mutual exchange of preliminary results for  the  $\eta  N$  S-wave
      scattering length prior to publication.

      We  are also indebted to Prof.  B.M.K.  Nefkens for pressing us
      very hard to apply  our  model  to  extracting  S-wave  scattering
      length,  and  to  all suggestions and criticism he made during the
      process\footnote{This  work  has  been  partly  supported  by   EC
      contract CI1$^{*}$-CT-91-0894.}.

\section*{Table captions}
\begin{description}
\item[Table 1.]
{\em Table 1.} Values of the S-wave scattering
      length for various models.
\end{description}
\section*{Figure captions}
 \begin{description}
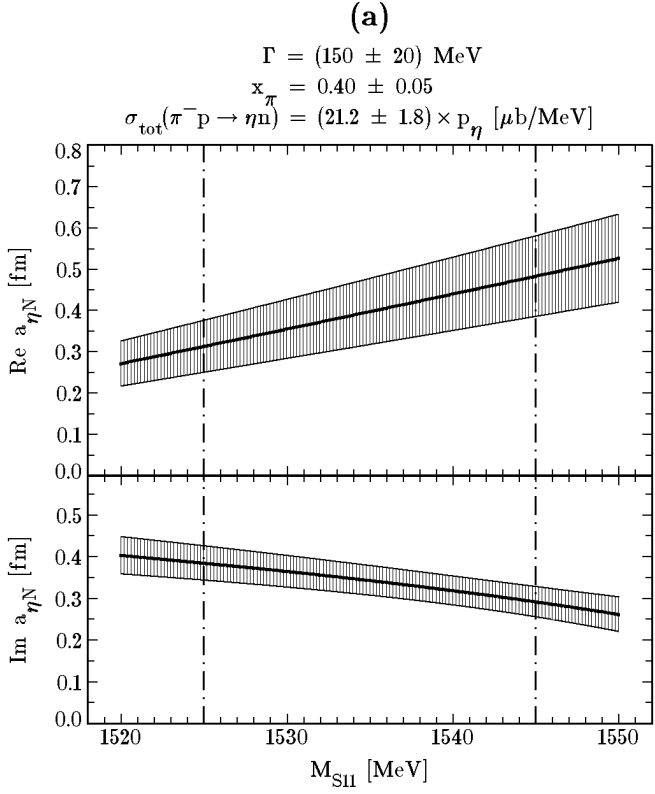
      \item[Fig.  1.]
      Dependence  of  the S-wave scattering length in a single resonance
      model on different input parameters:\\
      {\bf a.} Dependence  on  resonance  mass.   For  other  parameters
      influencing  $a_{\eta N}$ we have the following values:  resonance
      width  $\Gamma=(150\pm  20)$  MeV,   $\pi   N$   branching   ratio
      $x_\pi=(0.40  \pm  0.05)$  and  $\sigma^{tot}_{\pi^- p \rightarrow
      \eta n}/p_\eta =(21.2 \pm 1.8) \ \mu$b/MeV.  \\
      {\bf b.} Dependence on resonance width.  Values used for resonance
      mass  $M=(1535  \pm 10)$ MeV, resonance width $\Gamma=(150\pm 20)$
      MeV,  $\pi  N$  branching  ratio  $x_\pi=(0.40  \pm   0.05)$   and
      $\sigma^{tot}_{\pi^-  p \rightarrow \eta n}/p_\eta =(21.2 \pm 1.8)
      \ \mu$b/MeV.  \\
      {\bf c.} Dependence on $\pi N$ branching ratio.  Values  used  for
      resonance   mass   $M=(1535   \pm   10)$   MeV,   resonance  width
      $\Gamma=(150\pm 20)$ MeV and  $\sigma^{tot}_{\pi^-  p  \rightarrow
      \eta n}/p_\eta =(21.2 \pm 1.8) \ \mu$b/MeV.  \\
      {\bf  d.}  Dependence  on $\pi^- p \rightarrow \eta n$ total cross
      section near $\eta N$ threshold.  Values used for  resonance  mass
      $M=(1535  \pm  10)$  MeV, resonance width $\Gamma=(150\pm 20)$ MeV
      and $\pi N$ branching ratio $x_\pi=(0.40 \pm 0.05)$.
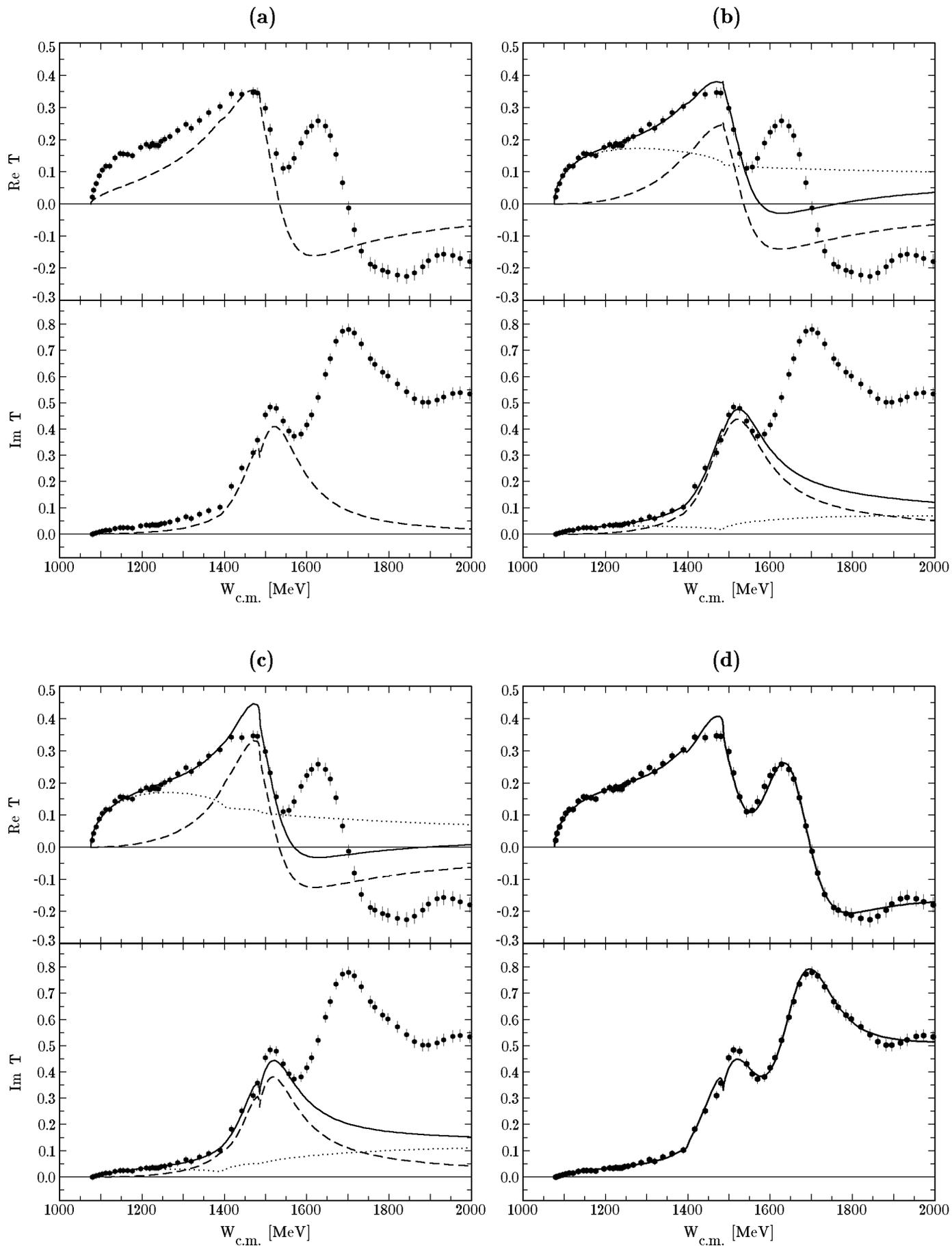
      \item[Fig.  2.] \hspace*{1.cm} \\
      {\bf a.}  Comparison  of  the  $\pi  N$  elastic  S-wave  T-matrix
      obtained  in  our  SR  model  (dashed  curve)  with the H\"{o}hler
      partial wave analysis (full dots) \cite{Hoe83}.  The used PWA does
      not  give  the  error  analyses for the partial wave T-matrices in
      \cite{Hoe83}, so the error bars given in the figure are defined in
      the  text  and reflect the statistical weight of the data set used
      in the minimization procedure.  \\
      {\bf b.} Comparision of the $\pi N$ elastic T-matrix (full  curve)
      obtained  in  our  SRBG  model  with  the  H\"{o}hler partial wave
      analysis (full dots) \cite{Hoe83}.  The  dashed  curve  represents
      the  resonance  contribution,  while  the  dotted  curve gives the
      background. \\
      {\bf c.} Comparision of the $\pi N$ elastic T-matrix (full  curve)
      obtained  in  our  CSRBG  model  with  the H\"{o}hler partial wave
      analysis \cite{Hoe83}.  The dashed curve represents the  resonance
      contribution, while the dotted curve gives the background. \\
      {\bf  d.}  Comparision of the $\pi N$ elastic T-matrix obtained in
      the  CCMRR  model  with  the  H\"{o}hler  partial  wave   analysis
      \cite{Hoe83}.
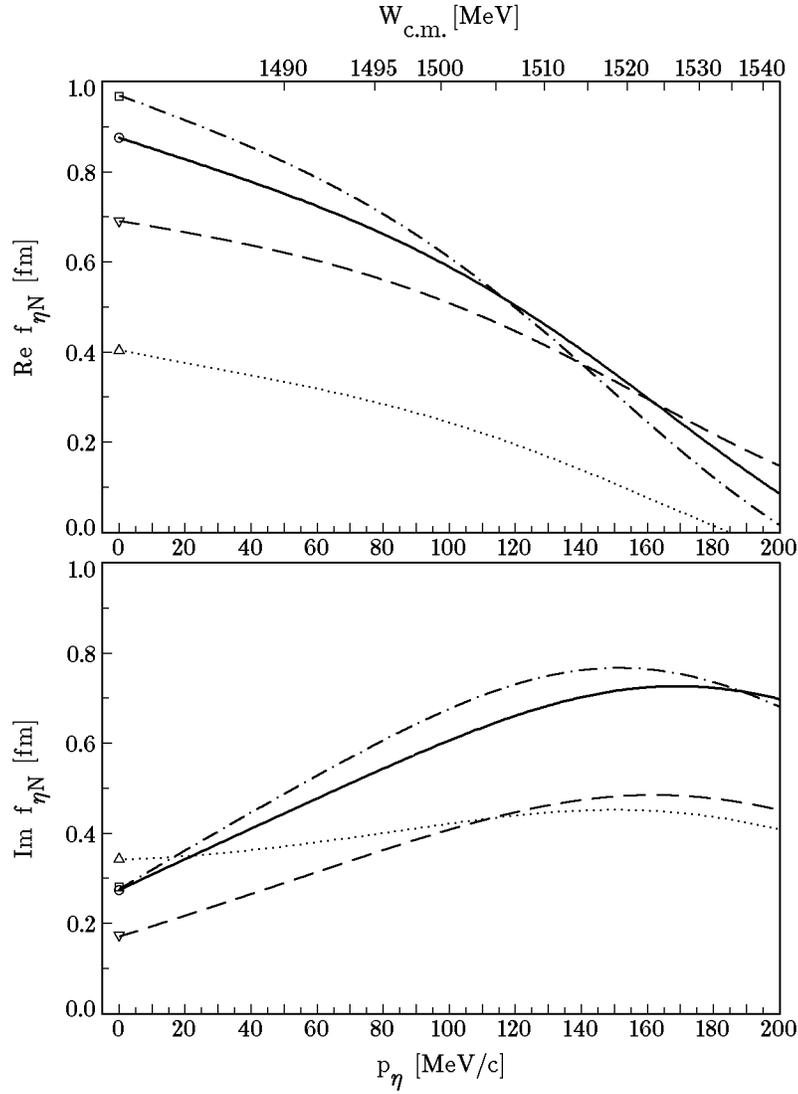
      \item[Fig.    3.]  The  dependence  of  the  $\eta  N$  scattering
      amplitude  upon  the  $\eta$  momentum.    The   dotted,   dashed,
      dash-dotted and full lines represent prediction of SR, SRBG, CSRBG
      and  CCMRU  models,  respectively.   Triangle,  inverse  triangle,
      square  and  open  circle  at  the y-axes show the $\eta N$ S-wave
      scattering length values obtained by  numerical  extrapolation  of
      corresponding scattering amplitudes.
\end{description}
\newpage\noindent
\begin{center}
      {\large \bf Table \vspace{1.cm} 1}  \\
\begin{tabular}{ccc}
\hline\hline
      &$Re\ a_{\eta N}$ [fm]&$Im\ a_{\eta N}$ [fm] \\
\hline\hline
      Bhalerao-Liu \cite{Bha85,Liu93} & 0.27 & 0.22 \\
      & 0.28 & 0.19 \\
\hline
      "modified"  Bhalerao-Liu & 0.38 & 0.31 \\
                               & 0.44 & 0.30 \\
\hline
      Arima et al.  \cite{Ari92} & 0.98 & 0.37 \\
\hline
      Wilkin \cite{Wil93} & 0.55 $\pm$ 0.20 & 0.30 \\
\hline
      Abaev and Nefkens \cite{Aba94} & 0.62 & 0.30 \\
\hline
      SR of this paper & 0.404 $\pm$ 0.117 & 0.343 $\pm$ 0.058 \\
\hline
      SRBG of this paper & 0.691 & 0.174 \\
\hline
      CSRBG of this paper & 0.968 & 0.281 \\
\hline
      \  \  CCMRU model with 3R in $P_{11}$\ \ &\ \ 0.886 $\pm$ 0.047\ \
      &\ \ 0.274 $\pm$ 0.039\ \ \\
      \ \ CCMRU model with 4R in $P_{11}$\ \ &\ \ 0.876 $\pm$  0.047\  \
      &\ \ 0.274 $\pm$ 0.039\ \ \\
\hline\hline
\end{tabular}
\end{center}

\newpage
\bigskip
\bigskip
\bigskip
\begin{thebibliography}{Sva 82a}
      \bibitem{Bha85}
          R.S. Bhalerao and L.C. Liu,
          Phys. Rev. Lett., {\bf 54}, 865 (1985)
      \bibitem{Bro79} R.M. Brown, A.G. Clark, P.J. Duke, W.M. Evans,
                R.J. Gray, E.S. Groves, R.J. Ott, H.R. Renshall,
                A.J. Shah,
                 J.J. Thresher and M.W. Tyrrell,
                 Nucl. Phys.,{\bf B153}, 89 (1979)
      \bibitem{Ari92}
          M. Arima, K. Shimizu and K. Yazaki,
          Nucl. Phys., {\bf A543}, 613 (1992)
      \bibitem{Wil93}
          C. Wilkin,
          Phys. Rev., {\bf C47}, R938 (1993)
      \bibitem{Aba94}
          V.V. Abaev and B.M.K. Nefkens,
          private communication
      \bibitem{Liu86} L.C. Liu, Q. Haider,Phys. Rev. {\bf C 34}, 1845
                (1986)
      \bibitem{Kru93} B.  Krusche, Proc. of the I TAPS Workshop,
      Alicante 1993 (World Scientific)
      \bibitem{Liu93}
          L.C. Liu,
          Acta Phys. Pol., {\bf B24}, 1545 (1993)
      \bibitem{Bat94}
      M.~Batini\'{c},  I.~\v{S}laus,  A.~\v{S}varc  and  B.M.K.~Nefkens,
      Rudjer  Bo\v{s}kovi\'{c}  Preprint  {\bf  IRB-FEP-01/95},  to   be
      published in Phys.  Rev.  C, available as paper nucl-th/9501011 at
      {\tt xxx.lanl.gov} or {\tt babbage.sissa.it}.
      \bibitem{Gol64} M.L. Goldberger and K.M. Watson: Collision theory,
          John Wiley \& Sons, Inc., New York - London - Sydney (1964)
      \bibitem{Tay72} J.R. Taylor, Scattering Theory: The Quantum
                 Theory on Nonrelativistic Collisions, John Wiley \&
                 Sons, Inc. New York, London, Sydney, Toronto (1972)
      \bibitem{Bin73}
           D.M.~Binnie et al.,
           Phys. Rev. {\bf D8}, 2793 (1973)
      \bibitem{Pdg92} Particle data group,
                    Phys.  Rev., {\bf D45} (1992)
      \bibitem{Pdg94} Particle data group,
                  Phys. Rev. {\bf D50}, 1177 (1994)
      \bibitem{Cla92} M. Clajus and B.M.K. Nefkens,
               in $\pi$N Newsletter,{\bf No 7}, ed. G. H\"{o}hler,
               W. Kluge and B.M.K. Nefkens, 76 (1992)
      \bibitem{Dei69} W. Deinet, H. M\"{u}ller, D. Schmitt,
                 H.M. Staudenmaier,
                 S. Buniatov and E. Zavattini,
                 Nucl. Phys., {\bf B11}, 495 (1969)
      \bibitem{Bul69} F. Bulos, R.E. Lanou, A.E. Pifer, A.M. Shapiro,
               C.A. Bordner, A.E. Brenner, M.E. Law, E.E. Ronat,
               F.D. Rudnick,
               K. Strauch, J.J. Szymanski, P. Bastien, B.B. Brabson,
               Y. Eisenberg, B.T. Feld, V.K. Kistiakowsky, I.A. Pless,
               L Rosenson, R.K. Yakamoto, G. Calvelli, F. Gasparini,
               L. Guriero, G.A. Salandin, A. Tomasin, L. Ventura,
               C. Voci
               and F. Waldner,
               Phys. Rev., {\bf 187}, 1827 (1969)
      \bibitem{Cut79} R.E. Cutkosky, R.E. Hendrick, J.W. Alcock,
                 Y.A. Chao, R.G. Lipes, J.C. Sandusky and R.L. Kelly,
                  Phys. Rev., {\bf D20}, 2804 (1979),
                  R.E. Cutkosky, C.P. Forsyth, R.E. Hendrick and
                 R.L. Kelly,
                 Phys. Rev., {\bf D20}, 2839 (1979)
      \bibitem{Hoe83} G. H\"{o}hler,
                {\em in} Landolt-B\"{o}rnstein, Elastic and Charge
                Exchange Scattering of Elementary Particles, Vol.
                {\bf  9}, Subvolume {\bf b}: Pion Nucleon Scattering,
                {\bf Part 2} (1983)
      \bibitem{AGS94} AGS PROPOSAL, UCLA-10-P25-224 (October 1994):
      ETA Production at Threshold in the Reactions \pieta and $K^- p
      \rightarrow \Lambda \eta$,
      George Washington University -- UCLA -- BNL -- Abilene Christian
      University -- Ru\dj er Bo\v{s}kovi\'{c} Institute -- JINR -- PNPI
      collaboration.
\end{thebibliography}
\end{document}